\documentclass[11pt]{article}
\usepackage{graphicx}
\usepackage{amssymb}
\usepackage{psfig}
\usepackage[dvips]{color}
\usepackage{sao2}
\def\beq{\begin{equation}}
\def\eeq{\end{equation}}
\def\etal{ et al.}
\def\m@th{\mathsurround=0pt }  % formula array number position centering 
\def\eqalign#1{\null\,\vcenter{\openup1\jot \m@th 
   \ialign{\strut\hfil$\displaystyle{##}$&$\displaystyle{{}##}$\hfil 
   \crcr#1\crcr}}\,} 
\begin{document}
\title
{Photometric study of fields of nearby pulsars \\
with the 6\,m telescope}

\author{
Kurt V.G.\inst{a}
\and Komarova V.N.\inst{b}
\and Fatkhullin T.A.\inst{b}
\and Sokolov V.V.\inst{b}
\and Koptsevich A.B.\inst{c}
\and Shibanov Yu.A.\inst{c}}

\institute {Astro Space Centre of RAS, Moscow \\
\and SAO RAS, Nizhnij Arkhyz\\
\and Ioffe Physical Technical Institute, St.Petersburg}
\maketitle

\begin{abstract}
Fields of Geminga, PSR B0950+08, PSR J1908+0734 and PSR J0108-1431
have been observed within the program of
optical investigation of nearby neutron stars carried out with the 6\,m
telescope. Our multicolour BVR$_c$I$_c$ photometry has yielded the following
magnitudes of Geminga: V=25$\fm$3$\pm$0.4; R$_c$=25$\fm$4$\pm$0.3.
The estimates of the 3$\sigma$ upper limits in the B and I bands 
are  25$\fm$4 and  24$\fm$9, respectively. This is in agreement
with the previous observations of Geminga. We suggest candidate
for optical counterpart of PSR B0950+08, with $R_c = 25\fm4\pm0.3$.
When compared to the HST data obtained in UV 
its flux shows that this old pulsar can be brighter in near-IR than
in near-UV. We have obtained for the first time images of the fields
of PSR J1908+0734 and PSR J0108-1431. Although no optical counterparts
have been detected, the magnitude upper limits obtained  suggest further
observations of these pulsars.
\end{abstract}

\section{Introduction}

\begin{table*}[t]
\caption{Some characteristics of the pulsars under investigation
}
\vspace*{0.3cm}
\hspace*{1.5cm}
\begin{tabular}{l|c|c|c|c}
\hline
\hline
&&&&\\
\multicolumn{1}{c|}{Pulsar}           &J0633+1746                 &B0950+08                   &  J0108-1134               & J1908+0734        \\
&&&&\\
\hline
&&&&\\
$\alpha_{2000}$                       & 6$^h$33$^m$54$\fs$02      & 9$^h$53$^m$9$\fs$32       & 1$^h$8$^m$8$\fs$20        & 19$^h$8$^m$17$\fs$01 \\
$\delta_{2000}$                       & +17$^o$46\arcmin11\farcs5 & +7$^o$55\arcmin35\farcs6  & $-14^o$31\arcmin46\arcsec & +7$^o$34\arcmin14\farcs36 \\
$D\!M$ (pc cm$^{-3})$                 & 2.9                       & 2.97                      & 1.83                      & 11.1 \\
$d$ (kpc)                             & $\sim 0.16$               & $\sim 0.13$               & $\sim0.09$                & $\sim0.58$        \\
$P$ (ms)                              & 237                       & 253                       & 807.6                     & 212.4            \\
$\dot P \times10^{-15}$ (s s$^{-1})$  & 11                        & 0.229                     & 0.82                      & 0.825             \\
$\tau$ (yrs)                          & $3.4\cdot 10^5$           & $1.74\cdot 10^7$          & $1.6\cdot 10^8$           & $4.07 \cdot 10^6$ \\
$\log B$ (G)                           & 12.20                     & 11.40                     & 12.40                     & 11.63 \\
$\log \dot E$ (erg s$^{-1})$           & 34.51                     & 32.75                     & 30.79                     & 33.53 \\
&&&&\\
\hline
\hline
\end{tabular}
\vspace*{0.3cm}

$D\!M$ -- dispersion measure, 
$d$ -- distance, $P$ -- rotational period,
characteristic age $\tau=P/(2\dot P)$,
magnetic field $B = 3.2 \times 10^{19}(P \dot P)^{1/2}$, 
energy loss $\dot E = 4\pi^2I\dot PP^{-3}$,  
where neutron star inertia momentum $I$ is accepted to be equal to $10^{45}$ g cm$^2$
\label{psr}
\end{table*}

In a vast list of more than 1000 radio pulsars there are but a few
objects detected in other wavelengths  ($\gamma-$rays, X-rays,
far-UV, optical, see e.g., Ulmer, 1998; Becker \&\ Tr\"{u}mper, 1999; 
Korpela E.J. \&\ Bowyer S., 1998; Mignani, 1998). Multiwavelength
observations bring a wealth of a new information on the radiation
mechanisms of pulsars which cannot be obtained in a narrow spectral band.
For young ($\la 10^4$ yrs) pulsars like those in the Crab and Vela
nebula the emission in the all spectral ranges is mainly of non-thermal
origin; it is produced by relativistic particles generated in
magnetospheres of rapidly rotating neutron stars (NSs).
Becoming older ($\ga 10^5$ yrs), pulsars rotate slower, 
the non-thermal component weakens and one can observe 
thermal emission from the surface of cooling NSs.
According to  standard NS cooling models
(e.g., Nomoto \&\ Tsuruta, 1987), at this age
their  surface temperature is about  $10^5$--$10^6$K 
and the maximum of the thermal emission 
lies in soft X-rays (0.1--2.4~keV) or in  far-UV.
Thermal emission with spectral temperatures in 
the above range has been detected at X-ray observations 
of some middle-aged radio pulsars (see, e.g., Becker \&\ Tr\"{u}mper, 1997).
Thermal X-ray emission was also detected  from several
radio silent objects identified as isolated neutron stars (INSs) 
(e.g., Neuh\"{a}user \& Tr\"{u}mper 1999). 

Simulations of INS cooling show that at certain conditions,
depending  on the NS mass/radius ratio, on equation of state and composition 
of superdense  matter in interiors and in surface layers of the star,  
(see, e.g., Yakovlev \etal\ 1999 for a recent review),
thermal evolution of a NS may strongly deviate from the standard model.  
Theoretical investigations combined with comprehensive  
observational studies of  thermal emission from 
radio pulsars and INSs of different ages enable one 
to understand which of the evolution scenarios are real. 
These studies are also of crucial importance for development 
of realistic models of NSs and deeper understanding  
of poorly known properties of superdense matter 
in their interiors.
 
Optical observations are an important part of the multiwavelength
studies of INSs and pulsars. They allow one to constrain the parameters
of the thermal emission in the Rayleigh-Jeans spectral region and
investigate the properties of the nonthermal radiation in optical bands.
Most pulsars and INSs, excepting the young Crab-pulsar and PSR 0540-69,
are faint optical objects. Thus, the multicolour photometry is the natural
first step to search for the pulsar optical counterparts and to obtain
information on their optical spectra. This is the main goal of the program
which is carried out at the 6\,m telescope during last several years.

The program includes a deep (up to 27$^m$) photometry of the fields
of some nearby pulsars. The most interesting results reported at present
are those obtained for the middle-aged ($\sim 10^5$ yrs) pulsar PSR B0656+14
(Kurt \etal, 1998;  Koptsevich \etal, 2000).
Deep multicolour photometry observations of the pulsar field taken with
the 6\,m telescope yielded the detection of the pulsar optical counterpart.
The pulsar magnitudes in BR$_c$I$_c$-bands were estimated for the
first time. The correspondent fluxes exceed the Rayleigh-Jeans extrapolation
of the thermal spectrum seen in the soft X-rays and EUV.

In this paper we report  photometry of four other nearby pulsars:
Geminga (J0633+1746), PSR B0950+08, PSR J1908+0734 and PSR J0108-1431.

Their parameters are presented in Table \ref{psr}. The data were taken
from the pulsar catalogue (Taylor \etal, 1995),
excepting Geminga's DM (Malofeev \&\ Malov, 1997).

Observations and data reduction are described in Section 2, Section 3 
summarizes the results, and some conclusions are given 
in Section 4.

\section{Observations and data reduction}

\begin{table*}[t]
\caption{Technical characteristics of CCD detectors}
\vspace*{0.3cm}
\hspace*{1.0cm}
\begin{tabular}{c|c|c|c|c|c}
\hline
\hline
  Detector  &   Size    &Pixel&     &   Gain    &Readout \\
  type      &(in pixels)&size &FOV &($e^-$/ADU)& noise\\
\hline
ISD017A     &$1060\times1170$&$0\farcs137$&$2\farcm5\times3\farcm0$&2.3&8    \\

Photometrics &$1024\times1024$&$0\farcs206$&$3\farcm5\times3\farcm5$&1.3&4  \\

 TK1034  &$1034\times1034$&$0\farcs207$&$3\farcm6\times3\farcm6$&1.2 &3\\
\hline
\end{tabular}
\label{ccd}
\end{table*}
\begin{table*}[t]
\caption{Observations of the PSR fields}
\vspace*{0.3cm}
\hspace*{1.0cm}
\begin{tabular}{c|c|c|c|c|c}
\hline
\hline
& &  &&Total  & \\
Date &Detector&Object& Filter & exposure, sec & Seeing\\
\hline
11.03.97&ISD017A&PSR J0633+1746& R$_c$  &   3$\times$600    &     $1\farcs6$\\
12.03.97&ISD017A&PSR J0633+1746& V      &       600         &     $1\farcs4$\\
13.03.97&ISD017A&PSR J0633+1746& B      &   5$\times$600    &     $1\farcs8$\\
19.01.99&Photometrics&PSR J0633+1746& I$_c$ &   2$\times$300    &$1\farcs0$\\
\hline
\hline
11.03.97 &ISD017A&PSR B0950+08  &  R$_c$ &  3$\times$600  & $1\farcs6$   \\
\hline
24.01.98 &ISD017A&PSR B0950+08  &  R$_c$ &  5$\times$600  & $2\farcs0$   \\
%            & &  R     &  4800 & $\sim 1\farcs7$\\
\hline
\hline
 9.07.99&Photometrics&PSR J1908+0734 &  V    &   3$\times$600     &   $0\farcs9$\\
 9.07.99&Photometrics&PSR J1908+0734 &  R$_c$&   2$\times$600     &   $1\farcs0$\\
 9.07.99&Photometrics&PSR J1908+0734 &  I$_c$&    100             &   $1\farcs0$\\
10.08.99&Photometrics&PSR J1908+0734 &  B    &   4$\times$600     &   $1\farcs4$\\
\hline
\hline
10.08.99 &TK1034&PSR J0108-1134& B    & 3$\times$300 &$1\farcs5$\\
10.08.99 &TK1034&PSR J0108-1134& V    & 2$\times$300 &$1\farcs5$\\
10.08.99 &TK1034&PSR J0108-1134& R$_c$& 5$\times$300 &$1\farcs8$\\
10.08.99 &TK1034&PSR J0108-1134& I$_c$& 5$\times$300 &$1\farcs6$\\

\hline
\hline
\end{tabular}
\label{log}
\end{table*}

Optical observations of the pulsar fields  were carried out with
the 6\,m telescope BTA on March, 1997, January, 1998, on January,
July and August, 1999 with a CCD detector mounted at the prime focus with
the set of filters close to the Johnson-Cousins system. 
Different CCD detectors were used. Some of their characteristics are
given in Table \ref{ccd}. Seeing varied from night to night between
$0\farcs9$ and $1\farcs8$.
Table \ref{log} gives details on each observing run.

Standard data reduction, including bias subtraction, account for 
dark current, correction for non-uniformity of the detector 
sensitivity (flat-fielding), and removing of cosmic ray events,
was performed making use of MIDAS procedures.
To compensate fringes we used the so-called superflats, i.e. flats obtained directly
from the sky by median-combining several science exposures taken during the night.
The individual flat-fielded exposures were stacked together, yielding
combined images of the fields under investigation.

For the photometric calibration we used  Landolt's standards (1992).
The absolute fluxes $F_j$ in [erg cm$^{-2}$s$^{-1}$Hz$^{-1}$] were
calculated using equations 
\beq 
\log F_j = -\left( 0.4 M_j + M_j^0\right),
\label{eq:fl_mag}
\eeq 
with the zero-points provided by Fukugita \etal (1995):
\beq
\eqalign{
M_B^0 = 19.396,\;\; M_V^0 = 19.445, \cr
M_R^0 = 19.520,\;\; M_I^0 = 19.623. \cr
        }
\label{eq:mag2fl_bta}
\eeq

Astrometric referencing of the images was performed using
the coordinates of selected field stars extracted from the USNO catalogue  
using the ESO Skycat tool. The final accuracy was about $1''$. 

\section{Results} 
\subsection{PSR J0633+1746} 
\begin{figure*}[t]
\vspace*{-0.5cm}
\hspace*{0.5cm}
\includegraphics[width=8.0cm,height=9.0cm,bb=180 255 434 535,clip]{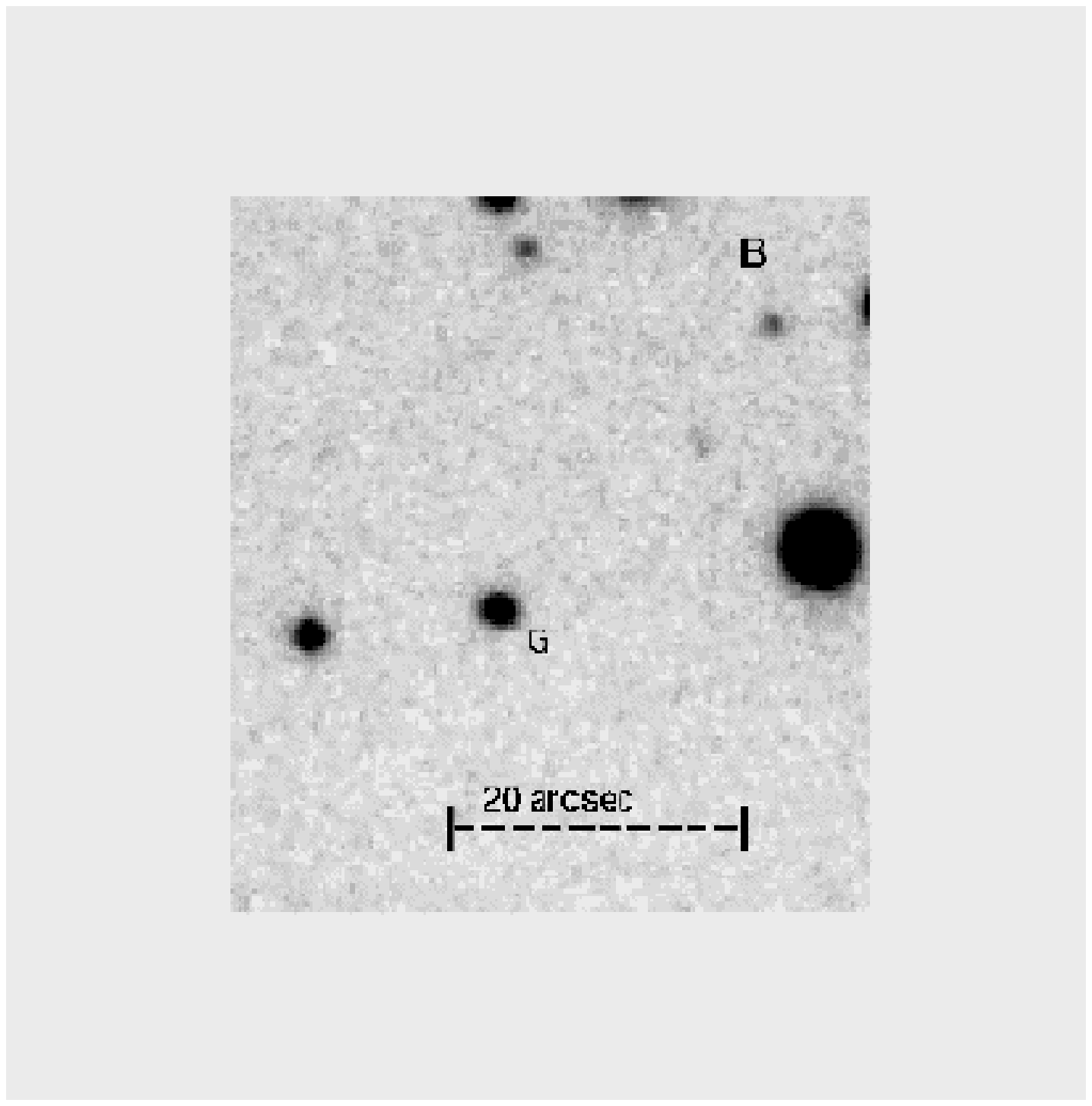}

\vspace*{-9.0cm}
\hspace*{8.8cm}
\includegraphics[width=8.0cm,height=9.0cm,bb=181 256 434 535,clip]{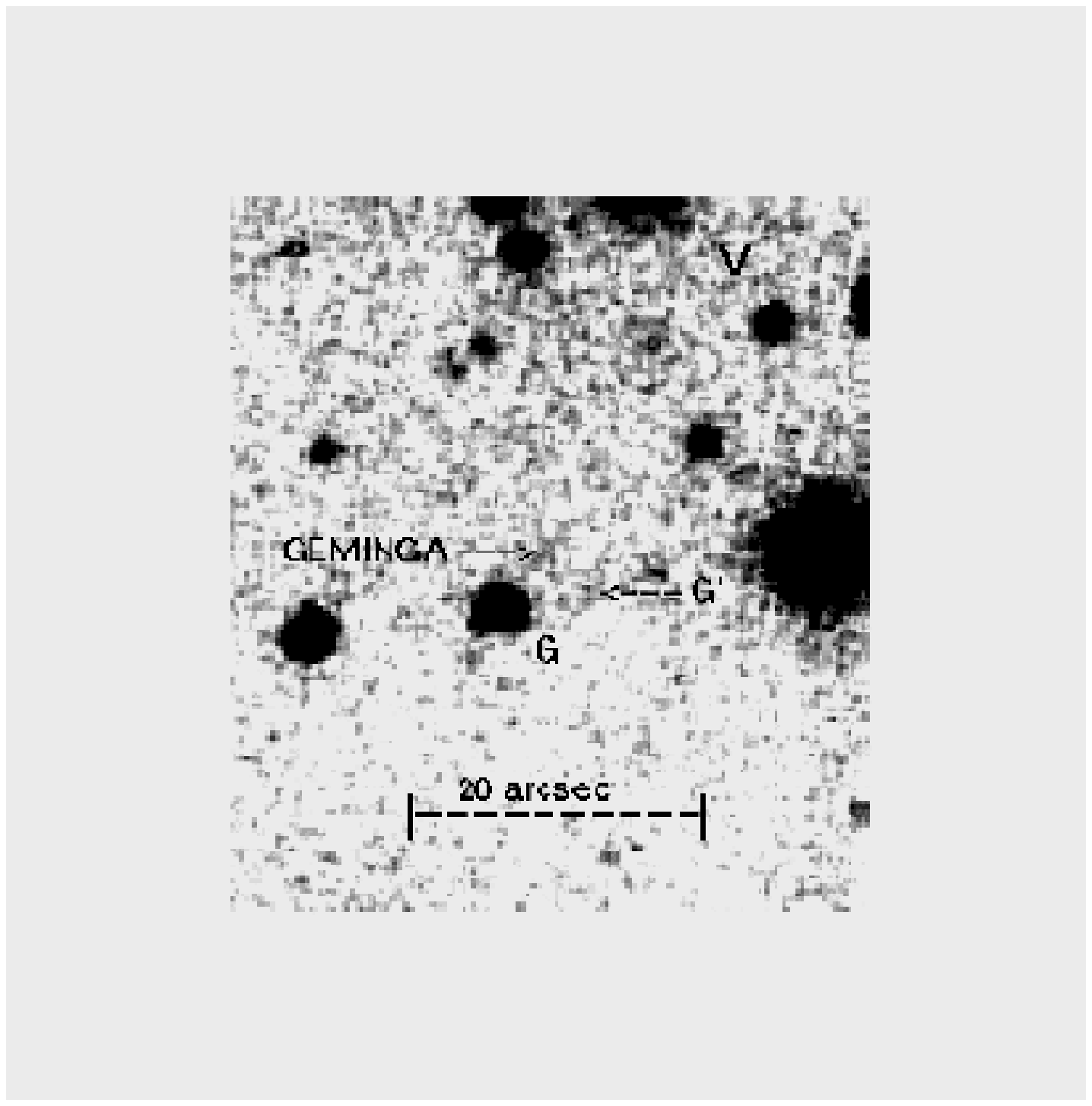}

\vspace*{0.1cm}
\hspace*{0.5cm}
\includegraphics[width=8.0cm,height=9.0cm,bb=176 256 440 535,clip]{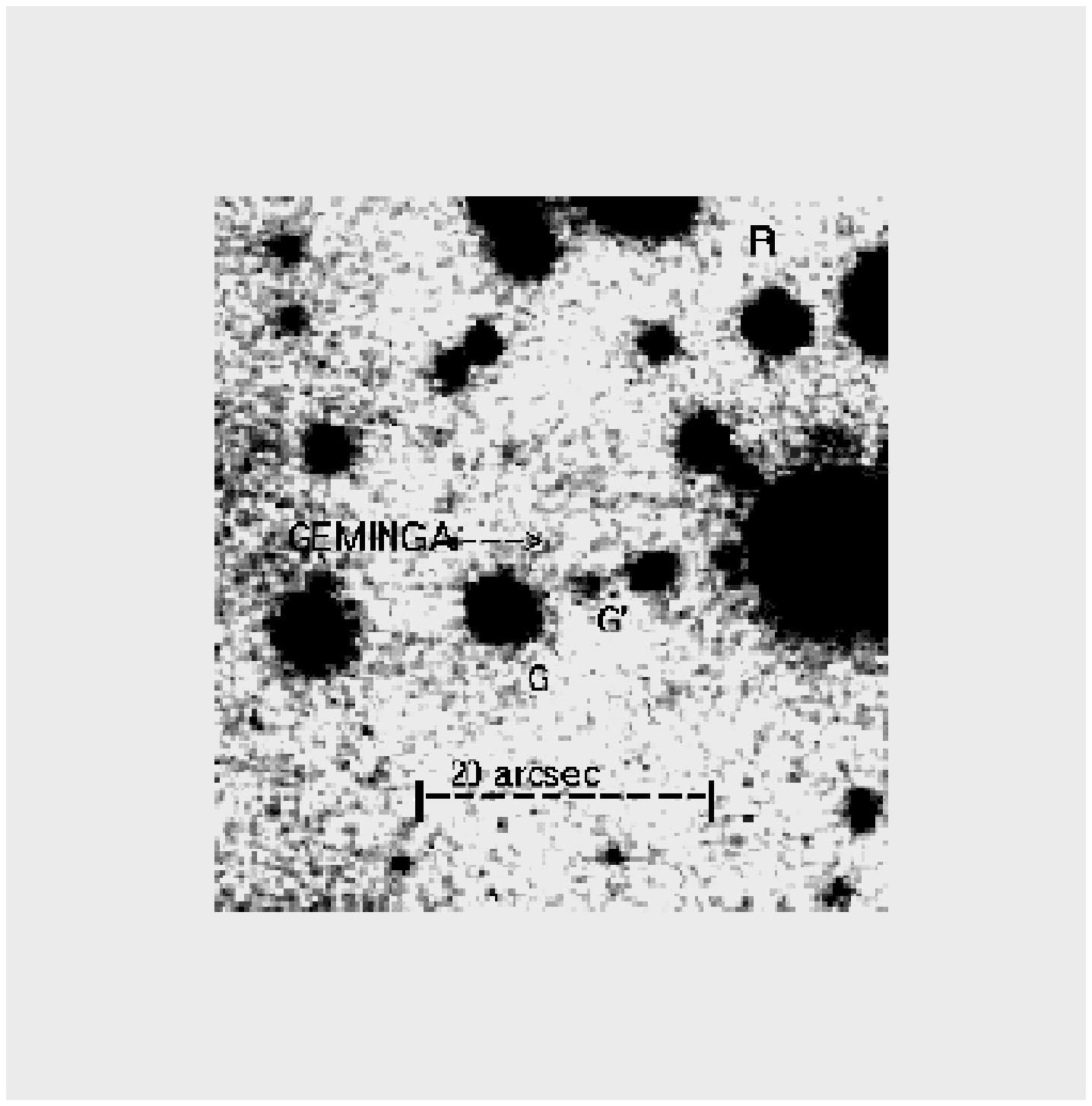}

\vspace*{-9.0cm}
\hspace*{8.8cm}
\includegraphics[width=8.0cm,height=9.0cm,bb=138 220 463 582,clip]{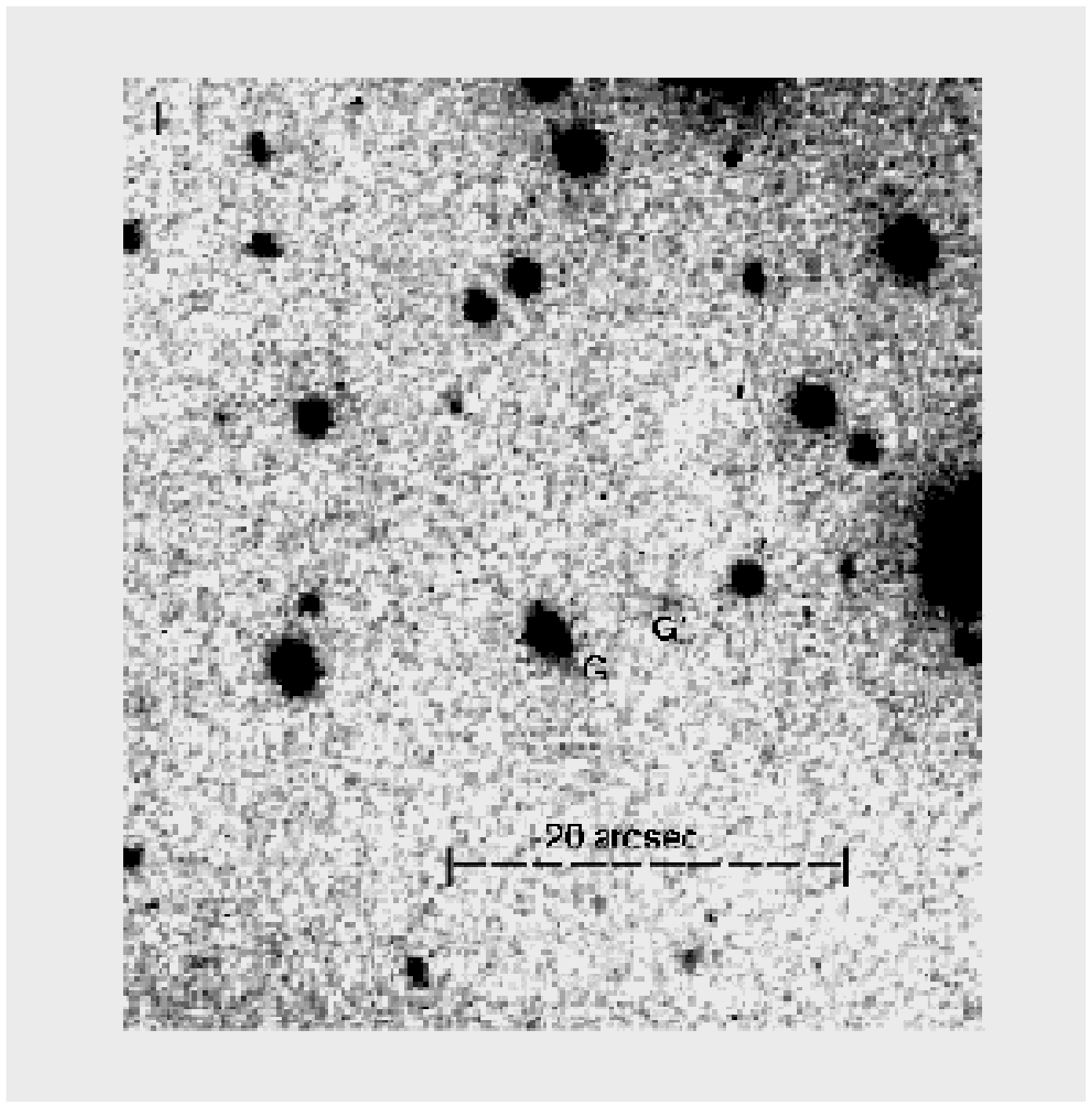}

\vspace*{-8.5cm}
\hspace*{15.2cm}
{\LARGE {\color{white} I}}

\vspace*{8.5cm}
\caption{The field of the pulsar Geminga (PSR J0633+1746)}
\label{gem2}
\end{figure*}

\begin{table*}[t]
\caption{Photometry of PSR J0633+17 (Geminga) and the nearest field objects}
\vspace*{0.3cm}
\hspace*{0.9cm}
\begin{tabular}{ccccccccc}
\hline
\hline
&\multicolumn{2}{c}{B}&\multicolumn{2}{c}{V}&\multicolumn{2}{c}{R}&\multicolumn{2}{c}{I}\\
&Magnitude&Flux$^a$&Magnitude&Flux&Magnitude&Flux&Magnitude&Flux\\
\hline
G       &22.1	&5.91	  &20.3$\pm$0.1&27.2&20.2$\pm$0.1&25.1&19.8$\pm$0.2&28.6\\
G$'$ &$>$25.4	&$<$0.278 &24.5$\pm$0.3&0.568&24.3$\pm$0.3&0.575&23.5$\pm$0.4&0.948\\
G$''^b$      & $>$25.4 & $<$0.278   & 25.3$\pm$0.4 & 0.272     & 25.4$\pm$0.3 & 0.209     & $>$24.9      & $<$0.256 \\
\hline
\end{tabular}
\vskip .3cm
\hspace*{0.9cm}
\begin{tabular}{ll}
$^a$    & Flux in $\mu$Jy \\
$^b$ & Geminga optical counterpart \\
\end{tabular}
\label{ph_gem}
\end{table*}
The middle-aged Geminga pulsar has been already detected
with the ground-based (CFHT, ESO 3.6\,m, NTT) and HST telescopes in different
bands and presently it is one of the well-studied pulsars in the the optical
domain (see, e.g., Bignami\ \&\ Caraveo, 1996, Mignani \etal, 1998).
It has been concluded that the optical emission of Geminga seems to be
thermal with a broad emission feature at 6000\AA.  Jacchia \etal (1999)
suggested a rough model to explain the observed excess of Geminga's emission
in the V-band as emission of hot ions in the strong magnetic field near
the NS surface. The results of Geminga's spectroscopy (Martin et al,
1998) with tentative detection of an apparent absorption feature at 6400\AA\
are generally consistent with the photometry.
In Figure~1 we present our images of the Geminga field in the B, V, R$_c$
bands taken on March, 1997 and in the I$_c$ band on January 19, 1999.
The pulsar position in the V and R images is indicated with arrow.
The objects suggested to be Geminga candidates at the first observations  
of this field in the optical range  (Halpern \&\ Tytler, 1988) are marked
with letters G and G$'$.
\begin{figure}[h]
\hspace*{0.1cm}
{\vbox{\psfig{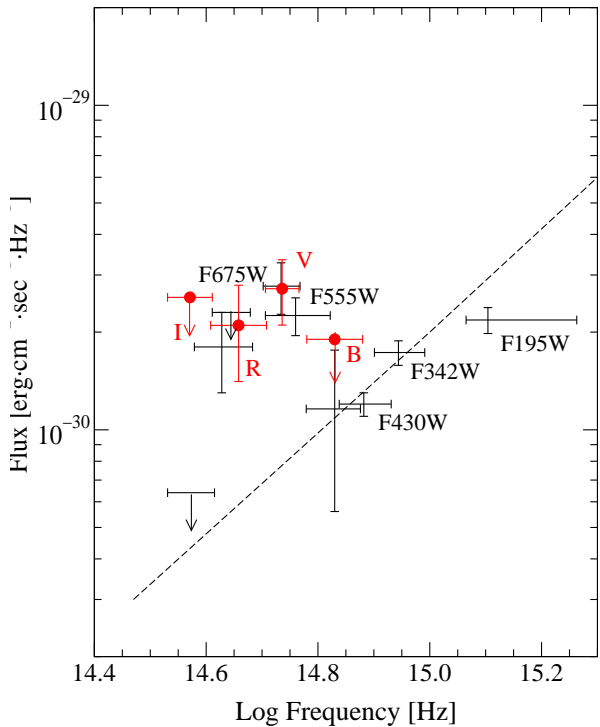}}}
\caption{
Summary of available data on the multiband HST and ground-based photometry
of Geminga. Data points are labeled according to the filters with
which they were obtained (see text for more explanation). Dashed line shows
the Rayleigh-Jeans part of the spectrum obtained at the blackbody fit of
the ROSAT data of Geminga for a neutron star with the 10~km radius
and the temperature of 5.77$\cdot 10^5$~K
(Halpern \& Ruderman 1993).
%  (Jacchia \etal, 1999)
}
\label{gem1}
\end{figure}
\par The photometry both of the pulsar optical counterpart (G$''$) and
the mentioned field  objects yields the magnitude estimates, which are
presented in Table \ref{ph_gem}. The obtained Geminga flux distribution
is in a good agreement with the data published so far. In Figure~\ref{gem1}
we present the broadband spectrum of Geminga based on all optical data
available. Filled circles correspond to our results, and arrows with bars
to the I and F675W upper limits. Our observations of the Geminga field in
the I-band were tentative and we could not get so deep images as had been
done before (Bignami \etal, 1996). Thus, the fading of Geminga's flux
at these wavelengths is still unconfirmed.

In all the optical bands (excluding the I one) the optical fluxes exceed
the Rayleigh-Jeans extrapolation of the thermal spectrum seen in the EUV
and soft X-rays (dashed line in Figure~\ref{gem1}). Such a spectrum
behaviour is  similar to that of the PSR B0656+14 and probably can be also
fitted by combination of both a magnetospheric and a thermal component.

\subsubsection{PSR B0950+08}
Like Geminga, the pulsar B0950+08 has been previously detected.
The field of this pulsar was observed with the HST in the UV-optical range
using the long-pass filter F130LP ($\lambda\lambda = 2310-4530$\AA\AA)
(Pavlov \etal, 1996). The only pointlike source was detected in the
image of $7\farcs4\times7\farcs4$ in size ($m_{F130LP}$ = 27.1; 
F = $0.051 \pm 0.003 \mu Jy$). Its location does not coincide with the radio
pulsar  position. The difference in positions is about $1\farcs85$, that may
be caused by systematic uncertainties in the HST Guide Star Catalogue (up to 
$1\arcsec$), which is used for HST astrometric referencing, 
as well as by uncertainties in the pulsar radio position ($0\farcs5$).

The field of this pulsar was observed with the 6\,m telescope in the R band
of the Cousins system on March, 1997 and  January, 1998.
\begin{figure}[t]

\includegraphics[width=8.0cm,bb=107 197 503 593,clip]{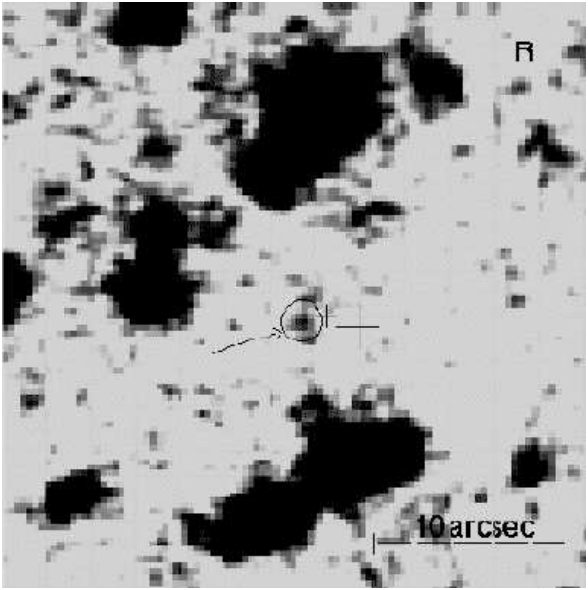}

\caption{The R-image of the PSR B0950+08 neighbourhood;
crosses indicate the pulsar radio position  and the position of the HST
UV-optical candidate with their uncertainties (small and big ones
respectively); arrow points to our candidate to optical pulsar
counterpart. The circle, which is $1''$ in radius, corresponds to
the accuracy of our astrometry.}
\label{psr0950}
\end{figure}
There are not enough reference stars from USNO catalogue in this field
(approximately $2\farcm5 \times 2\farcm5$ in size), that makes
an accurate astrometry impossible. To improve the situation we used
the image of this field ($9\arcmin \times 9\arcmin$) taken with
the Zeiss-600 telescope of SAO RAS. This enabled us to increase the accuracy of
the astrometrical referencing of the BTA images up to $1\arcsec$.
In Figure~\ref{psr0950} we present an image of PSR B0950+08 neighbourhood,
positions of the radio pulsar (small cross),
the UV-opical candidate (big cross) and the possible optical counterpart
(in circle) are indicated.
An object ($S/N  = 4.2$) has been found in the combined image, and its
position differs by $1\farcs5$ from the pulsar radio position. In case
the detected object is indeed the pulsar, this discrepancy may be due to
two reasons: different epochs of the radio and optical observations
with the lack of reliable data on the pulsar proper motion, and
uncertainties both in  our astrometry and the pulsar radio position.
The object magnitude is $R=25\fm5(3)$, corresponding flux is
$F_R = 0.19 \pm 0.04 \mu Jy$.
\begin{figure}[h]
\hspace*{0.1cm}
\includegraphics[height=10.0cm,bb=55 65 530 690,clip]{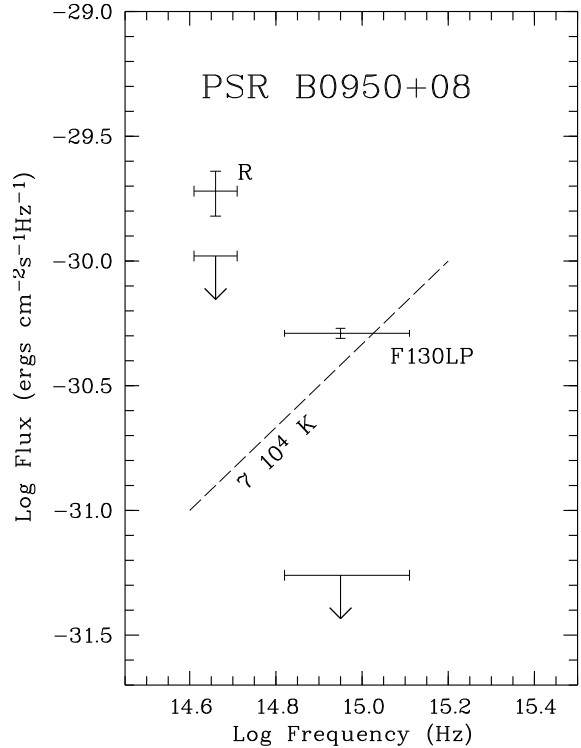}
\caption{Fluxes from PSR B0950+08 counterpart in the near-UV (HST) and
its possible optical candidate in R (6m telescope) ranges; arrows
with bars show the corresponding 3$\sigma$ upper levels on the flux of
undetected sources; dashed line reflects the blackbody spectral fit for
radiation from entire NS surface for the distance 
$d$ = 130~pc and NS radius of 13~km (Pavlov \etal, 1996).
}
\label{sp0950}
\end{figure}

Figure~\ref{sp0950} shows photometric data of the possible PSR B0950+08
counterpart based on the 6\,m telescope  (R-band) and HST (F130LP)
observations.

 If the observed object is the pulsar optical counterpart its
R-magnitude shows that old pulsars may be brighter in near-IR than in
near-UV region. This is in opposite to the expected spectral behaviour
and cannot be explained by thermal emission only. More observations
are needed.

\subsubsection{PSR J0108-1431}

\begin{figure*}[t]

\hspace*{0.5cm}
\includegraphics[height=9.25cm,bb=164 248 426 575,clip]{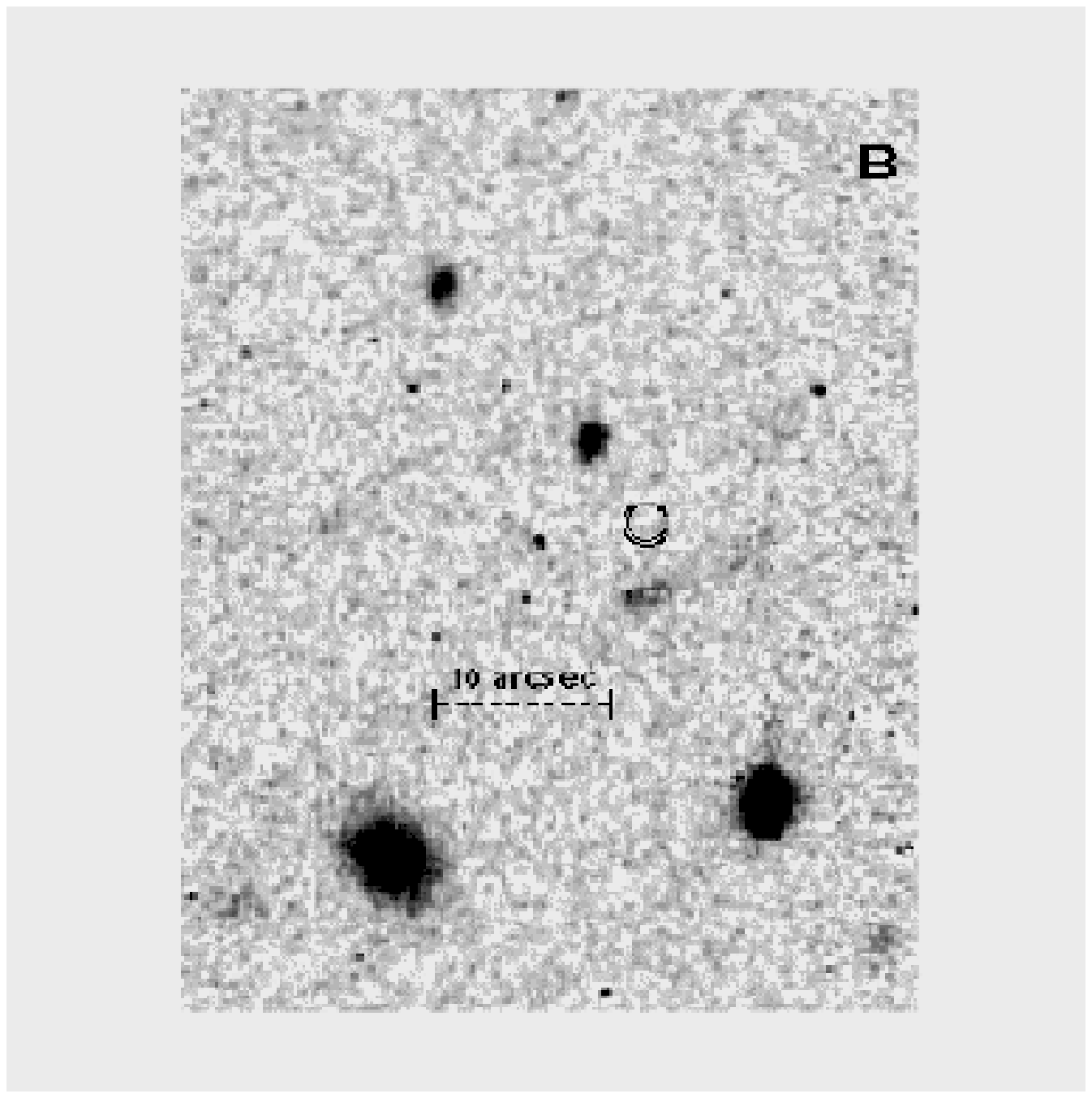}

\vspace*{-9.3cm}
\hspace*{8.2cm}
\includegraphics[height=9.25cm,bb=158 248 428 575,clip]{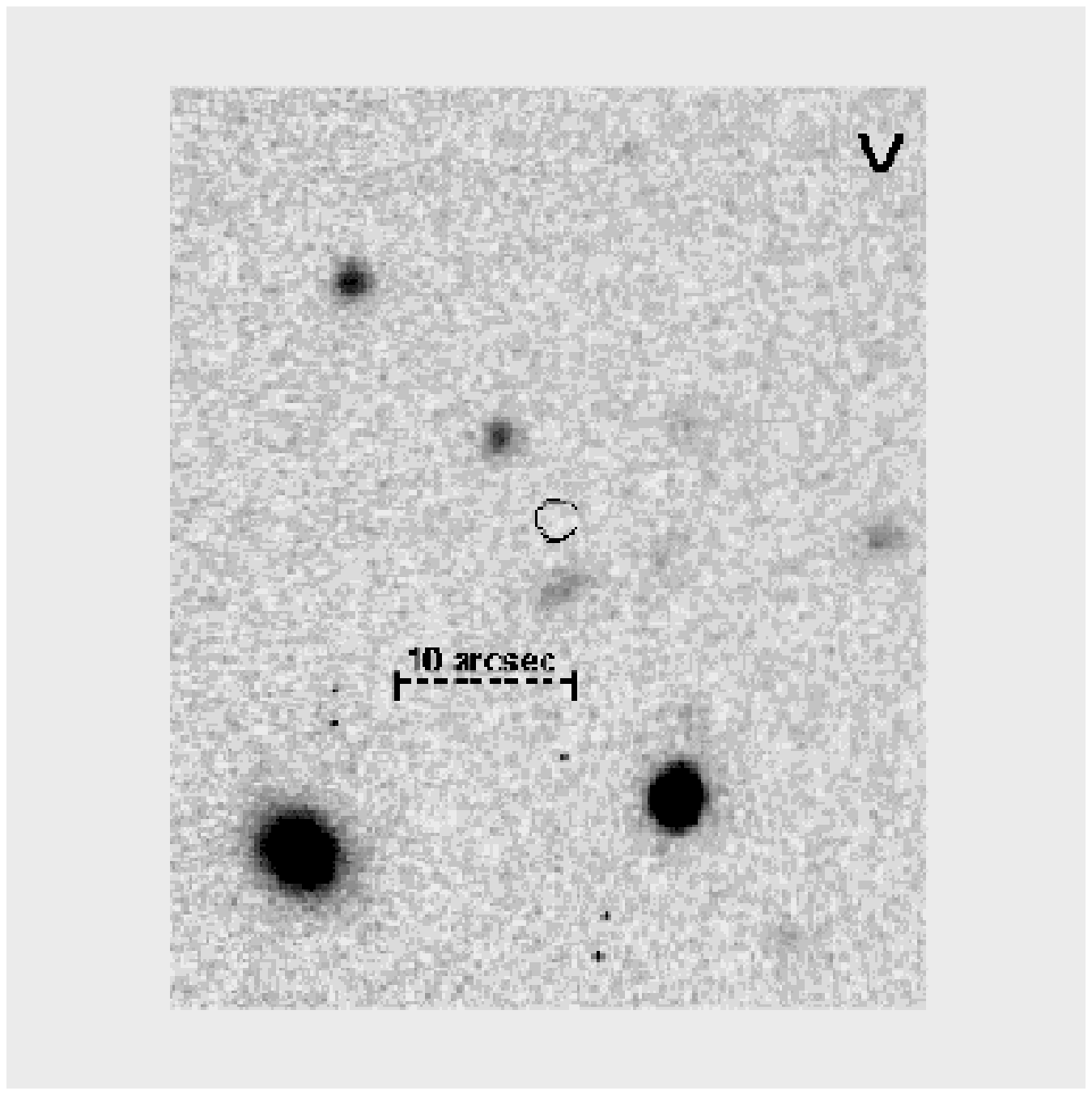}

\vspace*{0.2cm}
\hspace*{0.5cm}
\includegraphics[height=9.1cm,bb=158 248 428 575,clip]{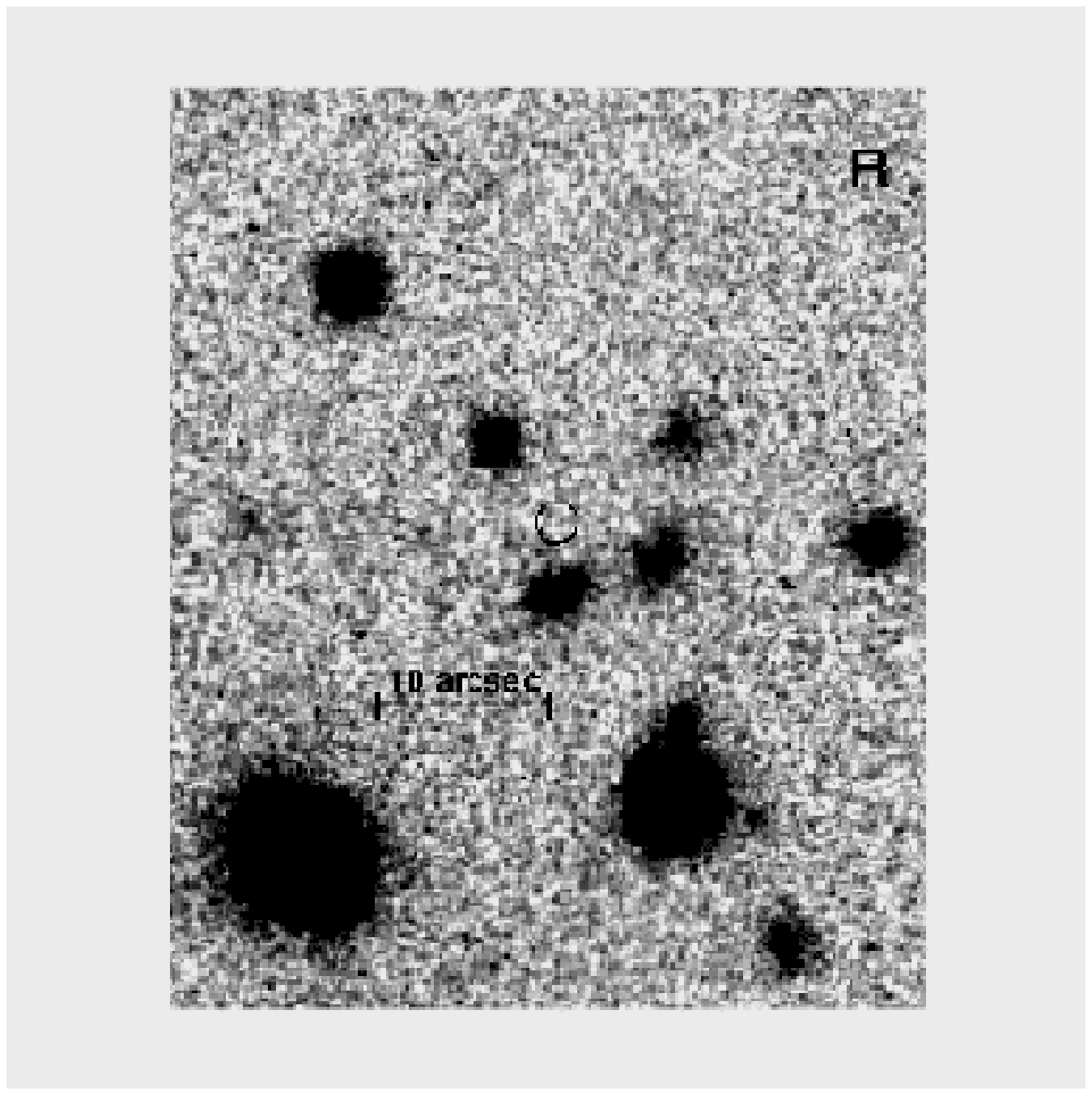}

\vspace*{-9.15cm}
\hspace*{8.2cm}
\includegraphics[height=9.1cm,bb=157 248 434 575,clip]{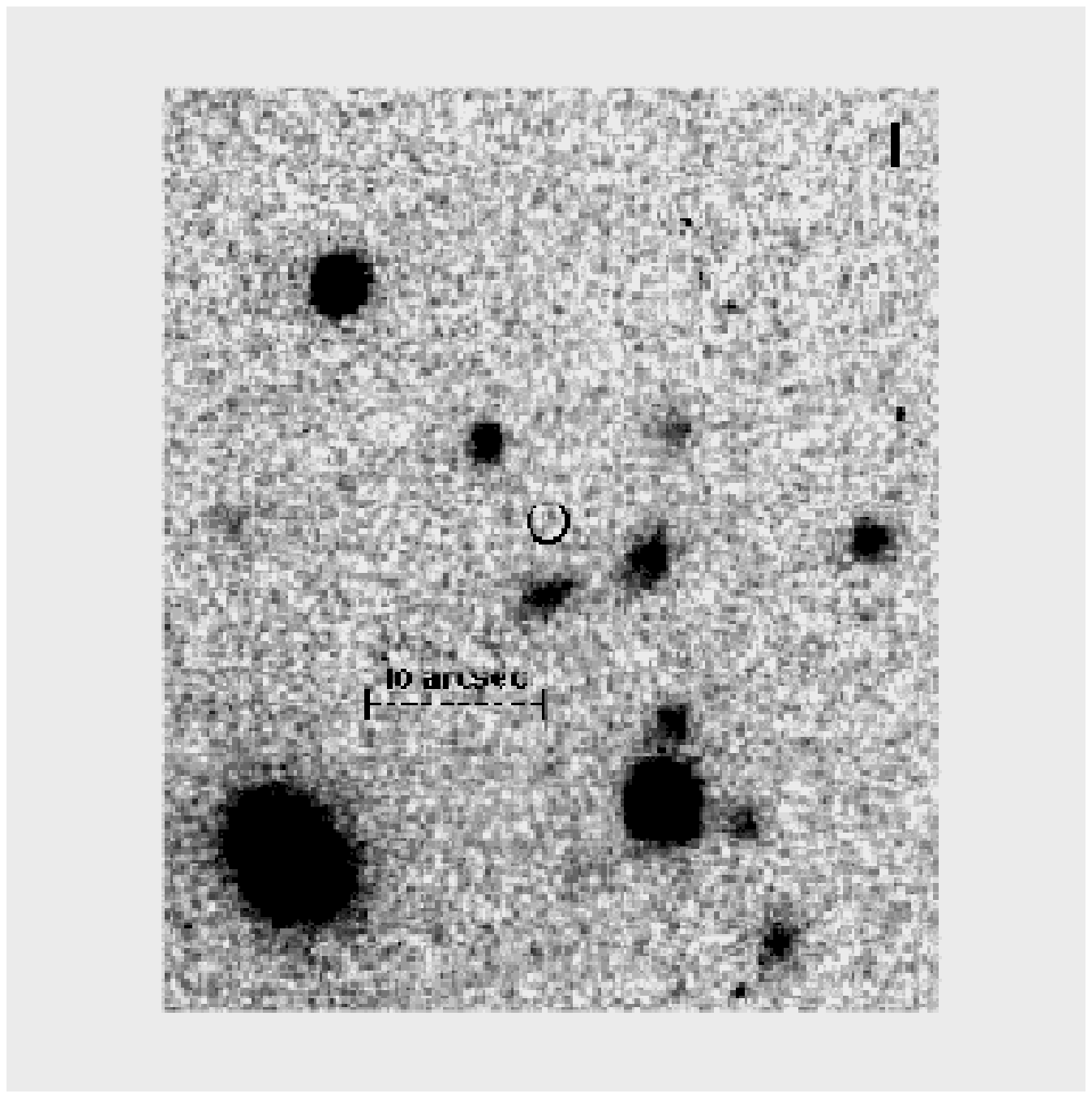}

\caption{The field of PSR J0108-1431, the pulsar radio position is marked
with the circle, its size corresponds to uncertainties in the astrometry}
\label{psr0108}
\end{figure*}

PSR J0108-1431 was discovered during a survey of the southern sky for 
pulsars using the Parkes 64\,m radio telescope (Tauris \etal, 1994).
In accordance with the galactic electron distribution density model 
(Taylor\ \&\ Cordes, 1993) the low dispersion measure suggests that
this pulsar is within 90~pc from the Sun.  Thus this is the closest
known neutron star, but it is quite old, $\tau\sim1.6\cdot 10^8$yrs.
The pulsar has not been detected in any high-frequency range.

\begin{figure*}[t]

\hspace*{0.2cm}
\includegraphics[width=8.0cm,bb=169 243 446 550,clip]{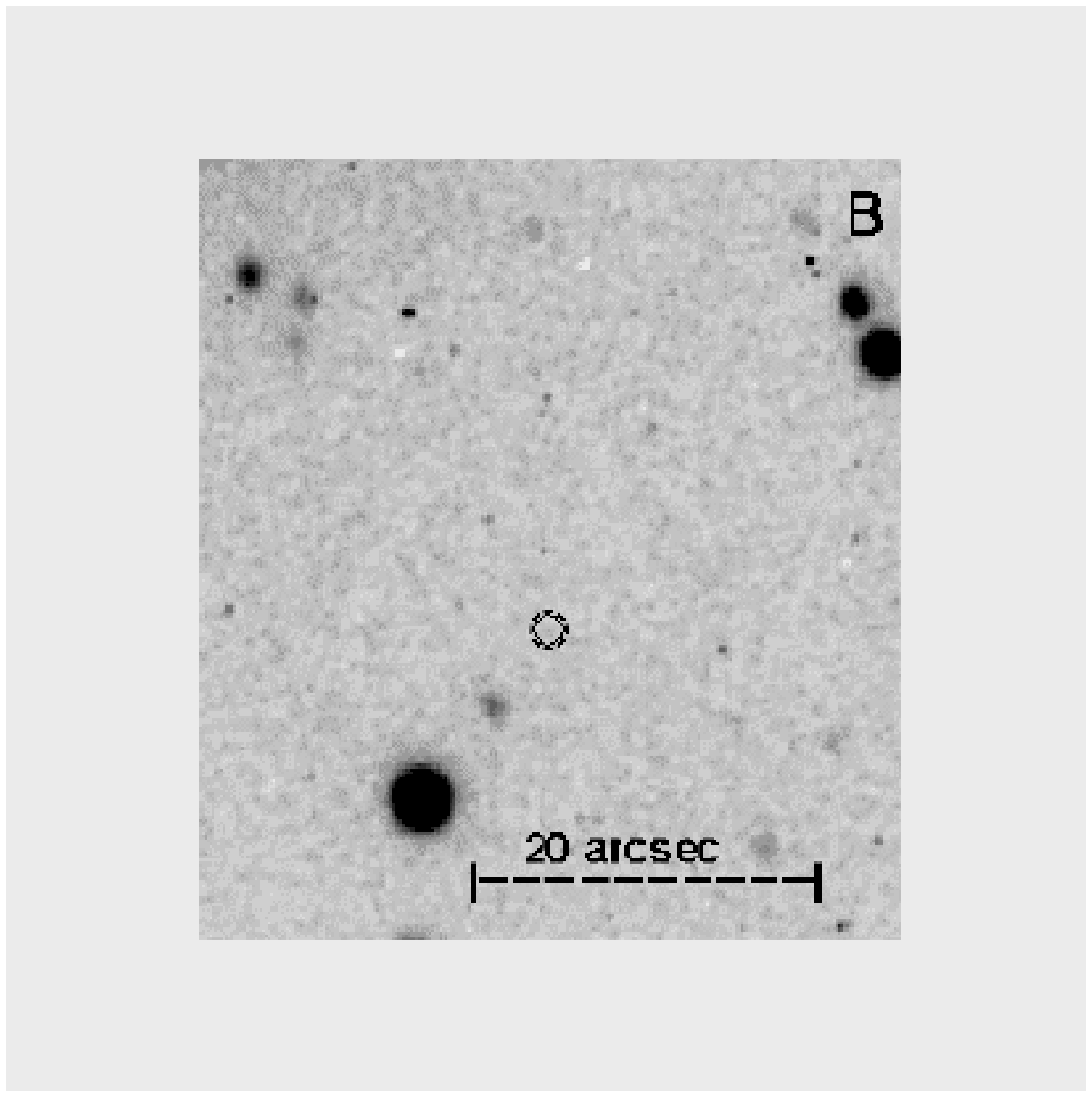}

\vspace*{-8.9cm}
\hspace*{8.4cm}
\includegraphics[width=8.0cm,bb=177 248 438 536,clip]{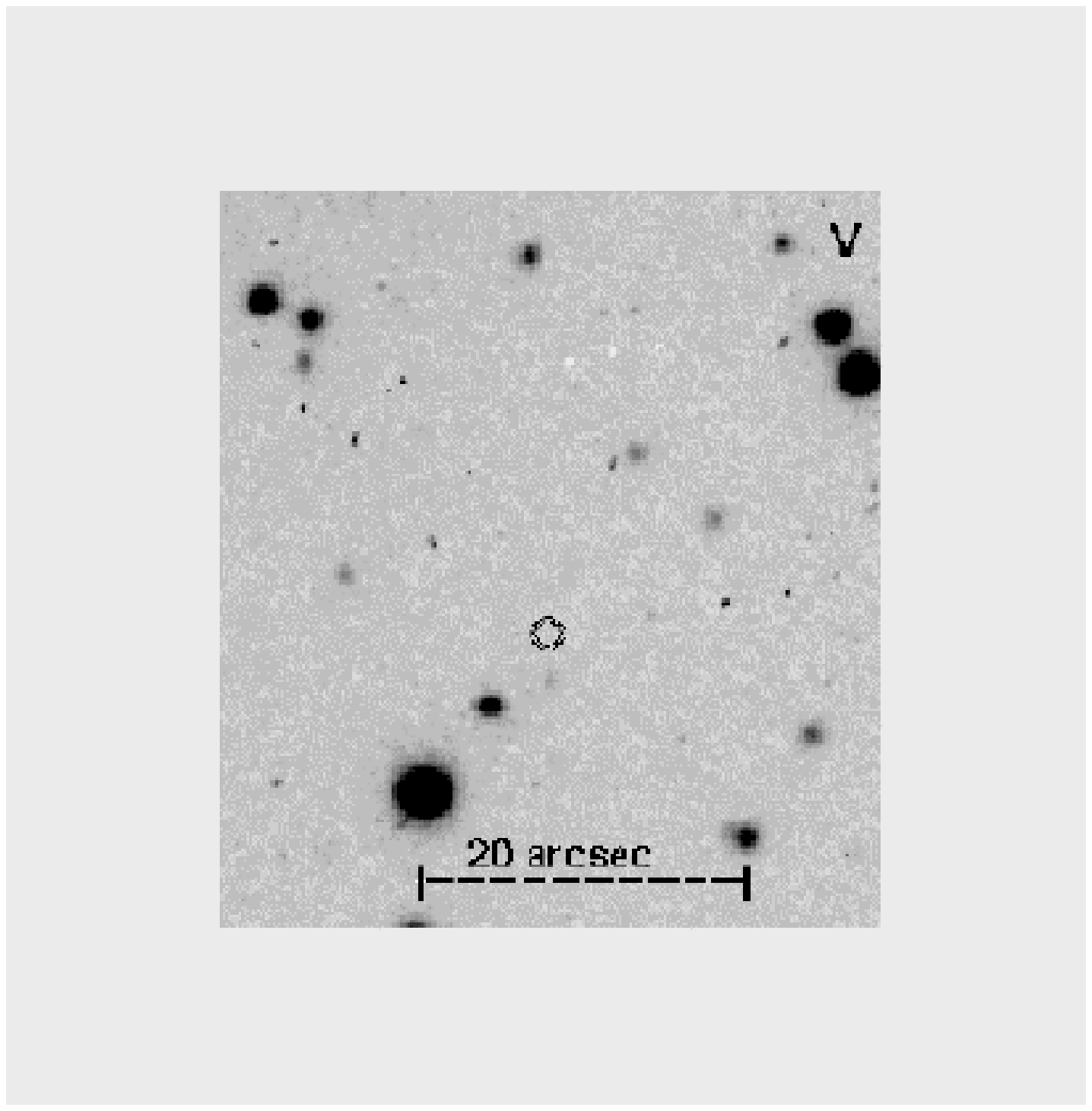}

\vspace*{0.2cm}
\hspace*{0.2cm}
\includegraphics[width=8.0cm,bb=177 248 438 537,clip]{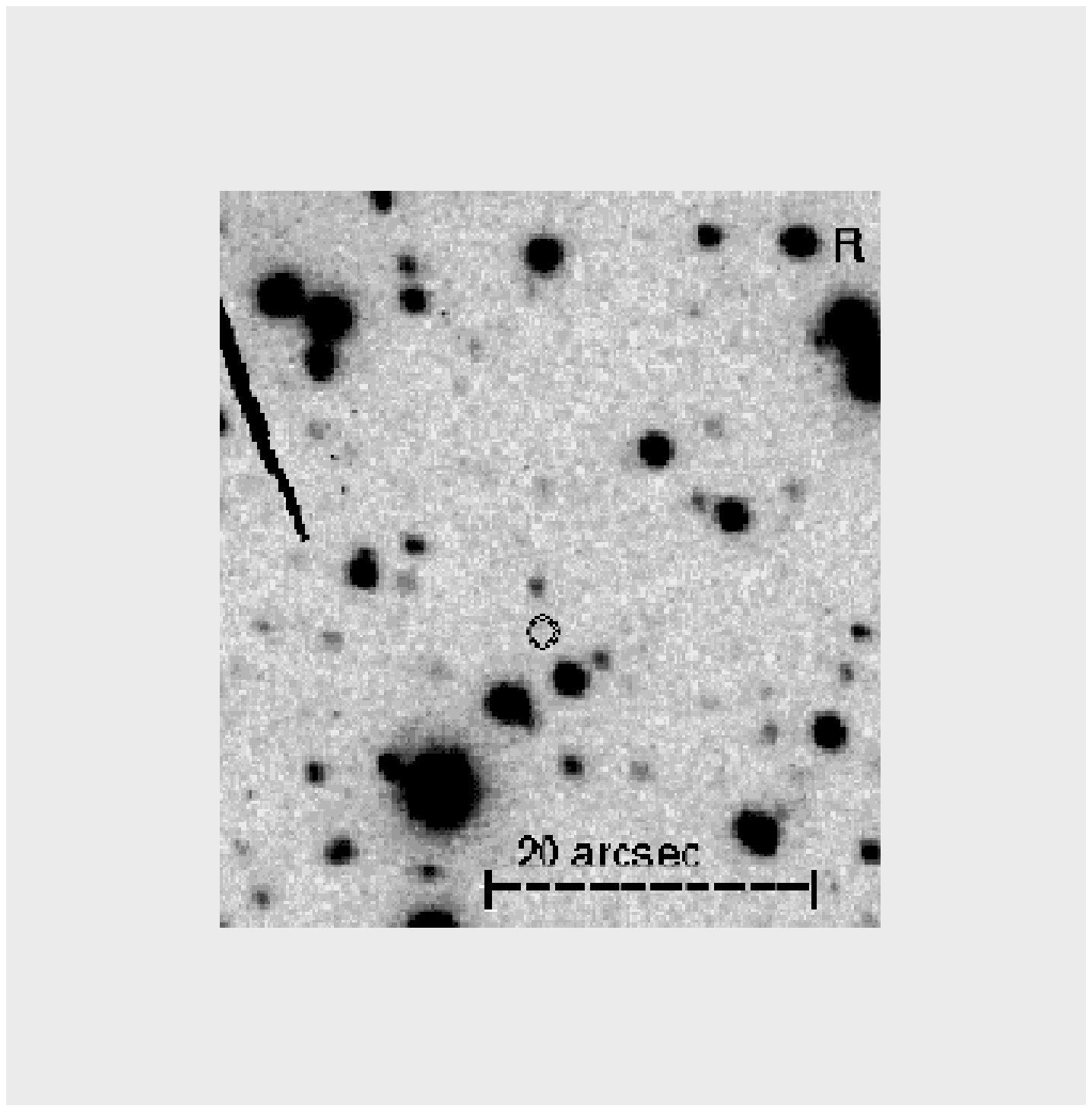}

\vspace*{-8.9cm}
\hspace*{8.4cm}
\includegraphics[width=7.9cm,bb=169 243 446 550,clip]{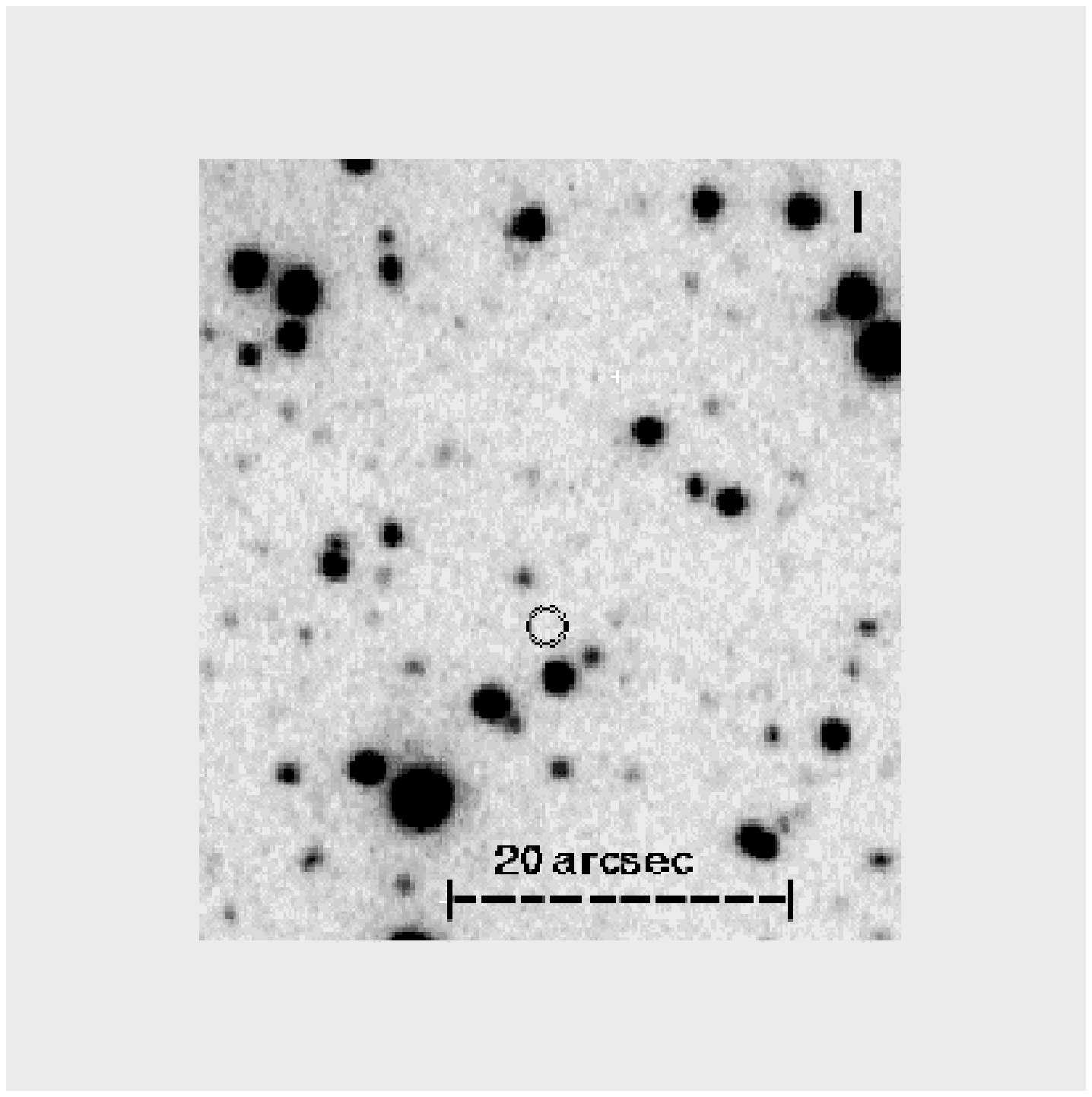}

\caption{The field of PSR J1908+0746; the circle marks the pulsar radio
position, its size corresponds to uncertainties in astrometry}
\label{psr1908}
\end{figure*}
Figure~\ref{psr0108} represents images of the PSR J0108-1431 neighbourhood
taken with the 6\,m telescope on August 1999. No object coinciding with
the radio pulsar in position was detected and we obtained the following
3$\sigma$-limits for the observed pulsar magnitudes: B $> 25\fm4$;
V $> 24\fm7$; R$_c$ $> 25\fm4$; I$_c$ $> 24\fm3$.
Further observations of this field might yield the detection of the INS
optical counterpart.

\subsubsection{PSR J1908+0746}

The pulsar PSR J1908+0746 was discovered at the Arecibo Observatory during 
the search for pulsars of low luminosity using the 305\,m radio telescope 
(Camilo\ \&\ Nice, 1995). For its characteristic age  this is old 
pulsar, while the rotational energy loss $\dot E$ is rather high.
In the absence of an unequivocal  theory  of the high-energy  
emission of pulsars, a high level of $\dot E/d^2$  is used as a rough 
indicator of likelihood that high-frequency emission from a pulsar 
can be detected (see, e.g., Goldoni\ \&\ Musso, 1996). 
Search for the emission in the high-frequency domain from pulsars 
with the highest values of $\dot E/d^2$ ($\sim$40 pulsars  
including PSR J1908+0746) yielded the detection of X-rays   
from 27 pulsars (Becker\ \&\ Trumper, 1997) and the $\gamma$-rays from 4
pulsars (Thompson \etal, 1994). No high-frequency emission has been
found so far from PSR J1908+0746 (Becker\ \&\ Trumper, 1997).
 
The field of this pulsar was observed in BVR$_c$I$_c$ bands
on July and August, 1999. The obtained images are shown in
Figure~\ref{psr1908}. The photometry yields the following 3$\sigma$
upper limits for the observed pulsar magnitudes:
B $> 26\fm0$, V $> 26\fm1$, R$_c$ $> 25\fm9$, I$_c$ $> 23\fm4$.
As this rather distant pulsar lies almost in the galactic plane,
its dereddened flux values can be derived only after an accurate
study of the interstellar extinction towards the pulsar position.

\section{Discussion and conclusions}

The deep photometric study of the PSR B0656+14 at the 6\,m telescope
has shown that the broadband spectrum of this middle-aged pulsar is
significantly of nonthermal origin (Kurt \etal, 1998; Koptsevich \etal,
2000). Its complicated shape differs from flat and featureless spectra
of young Crab-like pulsars and cannot be explained by a simple spectral model
(Pavlov \etal, 1997; Kurt \etal, 1998; Koptsevich \etal, 2000). Similar
behaviour, confirmed by our observations, shows the spectrum of the slightly
older pulsar Geminga. Some doubt still remains how deep is the fall of
Geminga's flux redward of R and whether its depth is restricted by the
emission level of the thermal component. More observations, including
the IR-range, are needed to address this question.
Nevertheless, current stage of  the optical studies allows us to suggest
that  the optical spectra of these
middle-aged NSs are likely to be very different from those of young ones
and may hint the spectral evolution of the optical emission with the pulsar
age.
 
The detection of the optical emission from the old pulsar PSR B0950+08,
if proposed counterpart candidate is confirmed by further observations,
may contribute to the above idea. The apparent excess
in the R-band may indicate the presence of spectral feature at these
wavelengths and raise a question whether thermal component is dominant
in the optical emission of old neutron stars ($>10^6$ yrs), which
is expected from the scenario of INSs evolution. Thus, search for optical
counterparts to old pulsars and cooling INSs detected in X-rays may be
of great importance.

We have observed the fields of the pulsars J0108-1431
and J1908+0734 for the first time. Although no optical counterparts
have been found, the estimated  magnitude upper limits suggest deeper 
observations of the fields of these promising objects. However,
the optical emission of PSR J1908+0734 seems unlikely to be detected
in the B and V bands due to extinction.

The search for optical emission of more INSs with the subsequent multicolour
photometric study is needed to shed light on the mechanisms of their
optical emission, to put constraints to the parameters of both thermal and
nonthermal components, to enrich our knowledge of the NS interior  and draw
a real picture of the neutron star evolution.

\begin{acknowledgements}
The work on pulsar study was partially supported by grant 1.2.6.4 
of Program "Astronomia", by INTAS (grant 96--0542) and RFBR (grant
99-02-18099). The authors are grateful to S.V.~Zharikov, N.A.~Tikhonov,
A.V.~Moiseev and V.R.~Amirkhanyan for assistance in observations.
VNK thanks A.I.~Kopylov, N.A.~Tikhonov and I.O.~Drozdovsky
for fruitful discussions and valuable comments.
\end{acknowledgements}

\end{document}